# Pronounced orbital-selective electron–electron correlation and electron–phonon coupling in V$_2$Se$_2$O

Mingzhe Hu[1,2,*], Ziyin Song[1,2,*], Jingwen Cheng[1,2,*], Gexing Qu[1,2], Zhanghuan Li[1,2], Yu Huang[1,2], Jundong Zhu[1,2], Guangyu Zhang[1,2,3], Dacheng Tian[1,2], Lan Chen[1,2], Zhijun Tu[4,5], Hechang Lei[4,5], Xiaoping Ma[1,2], Huaixin Yang[1,2], Zhongxu Wei[1], Genfu Chen[1,2,3,†], Hongming Weng[1,2,3,‡], Tian Qian[1,§], Hang Li[1,¶]

[1]*Beijing National Laboratory for Condensed Matter Physics and Institute of Physics, Chinese Academy of Sciences, Beijing 100190, China*

[2]*University of Chinese Academy of Sciences, Beijing 100049, China*

[3]*Songshan Lake Materials Laboratory, Dongguan 523808, China*

[4]*School of Physics and Beijing Key Laboratory of Optoelectronic Functional Materials & MicroNano Devices, Renmin University of China, Beijing 100872, China*

[5]*Key Laboratory of Quantum State Construction and Manipulation (Ministry of Education), Renmin University of China, Beijing 100872, China*

[*]These authors contributed equally to this work.

Corresponding authors: [†]gfchen@iphy.ac.cn, [‡]hmweng@iphy.ac.cn, [§]tqian@iphy.ac.cn, [¶]hang.li@iphy.ac.cn

**Orbital-selective many-body effects, in which electrons occupying different orbitals experience distinct interaction strengths, play a crucial role in correlated multiorbital materials. However, these effects usually manifest in a complex manner, obscuring their microscopic origins. Here, by combining angle-resolved photoemission spectroscopy measurements with theoretical calculations, we reveal pronounced orbital selectivity in both electron–electron correlation and electron–phonon coupling in the van der Waals material V$_2$Se$_2$O. Electron correlation induces distinct bandwidth renormalization exclusively in the V $d_{xy}$-derived band, while the bands mainly composed of the other $d$ orbitals remain essentially unrenormalized. Orbital-resolved analyses identify that the filling number and the bandwidth are decisive factors governing orbital-dependent correlation. Simultaneously, the $d_{xz/yz}$-derived band exhibits a sharp kink anomaly, arising from enhanced coupling to high-energy phonon modes dominated by oxygen vibrations. Such pronounced orbital selectivity positions V$_2$Se$_2$O as a rare and prototypical platform for unravelling the microscopic mechanisms of orbital-selective electron–electron and electron–phonon interactions, and offers guiding principles for the design of correlated multiorbital materials.**

Many-body interactions in condensed matter systems underlie a wide range of emergent quantum phenomena, including superconductivity[1–6], metal–insulator transitions[7–11], and density waves[12–21]. In multiorbital materials, such interactions—notably electron–electron correlation and electron–boson coupling—often operate in an orbital-dependent manner, yielding varying degrees of quasiparticle renormalization[22–32]. A canonical example is the iron-based superconductors, where angle-resolved photoemission spectroscopy (ARPES) reveals more pronounced narrowing of the $d_{xy}$ band than the $d_{xz/yz}$ bands, even leading to an orbital-selective Mott transition in iron chalcogenides, along with orbital-selective coupling to bosonic modes[25,26,29,33–38]. Similar behaviour occurs in ruthenates, which exhibit differential correlation strengths across the $t_{2g}$ orbitals, orbital-selective Mott physics, and orbital-specific dispersion kinks[23,27,30,39–44]. More recently, nickelate superconductors and kagome metals have also been found to host orbital-dependent bandwidth renormalization and kink anomalies[31,32,45,46]. These observations establish orbital-dependent interactions as a ubiquitous characteristic of correlated multiorbital systems, profoundly influencing emergent quantum phenomena such as unconventional superconductivity and non-Fermi-liquid behaviour.

Despite their prevalence, the microscopic origins of orbital selectivity remain intensely debated. Theoretical studies highlight the importance of Hund's coupling and crystal-field splitting in suppressing orbital fluctuations and promoting divergent interaction strengths across orbitals[27,47–49]. Orbital-selective electron correlation has been attributed to orbital filling, orbital degeneracy, bandwidth, van Hove singularity, and structural distortion[23,25,49–52], while orbital-selective electron–boson coupling has been associated with spin fluctuation, phonons, and electron correlation[53–55]. However, in many multiorbital systems, orbital selectivity does not emerge as a simple isolated effect, but is entangled with various competing interactions and often involves multiple intertwined microscopic factors. These complexities have long hindered efforts to establish a direct correspondence between orbital-dependent phenomena and their underlying origins.

Here we identify the van der Waals material V$_2$Se$_2$O as a model system that disentangles these complexities. Combining ARPES measurements with theoretical calculations, we clearly uncover two distinct orbital-selective signatures: pronounced bandwidth renormalization in the $d_{xy}$ band and a sharp kink anomaly in the $d_{xz/yz}$ band. The correlation-induced renormalization is quantitatively reproduced by local density approximation (LDA) + Gutzwiller calculations. Orbital-resolved density of states and occupancy analyses trace the correlation effects to the $d_{xy}$ orbital's narrow bandwidth and near-half-filled occupancy. The kink feature is consistent with electron–phonon coupling calculations, which attribute its orbital selectivity to strong coupling between in-plane oxygen vibrations and the $d_{xz/yz}$ orbital wavefunctions. These clear correspondences between experimental signatures and microscopic characteristics establish V$_2$Se$_2$O as a uniquely clean platform for elucidating the mechanisms that drive orbital differentiation in correlated quantum materials.

## Layered crystal structure

Figure 1a presents the crystal structure of V$_2$Se$_2$O, which adopts a layered structure with space group $I4/mmm$[56,57]. The structure consists of Se–V$_2$O–Se sandwich-like block layers, where V and O atoms form V$_2$O square-net planes, with Se atoms situated directly above and below the centre of each square. To experimentally confirm this arrangement, we performed scanning tunnelling microscopy (STM) and scanning transmission electron microscopy (STEM) measurements. The atomically resolved STM image of the cleaved (001) surface (Fig. 1b) reveals a uniform square lattice of Se atoms with a periodicity of ~3.9 Å, consistent with the in-plane lattice constant determined by X-ray diffraction (Extended Data Table 1). The cross-sectional STEM image viewed along the [100] zone axis (Fig. 1c) corroborates the staggered stacking of the Se–V$_2$O–Se layers along the $c$-axis, consistent with the structural model derived from X-ray diffraction refinement. The crystals can be mechanically exfoliated into large-area thin flakes with thickness down to ~3 nm (Fig. 1d), indicating weak van der Waals interlayer bonding. This mechanical exfoliability, reminiscent of established two-dimensional materials such as graphene and transition metal dichalcogenides, underlines the potential of V$_2$Se$_2$O as a promising platform for fundamental studies in two-dimensional electronics and nanoscale device applications.

## Multiband electronic structure

To elucidate the electronic structure of V$_2$Se$_2$O, we performed ARPES measurements complemented by density functional theory (DFT) calculations. Photon-energy-dependent ARPES spectra show minimal dispersion along the out-of-plane momentum direction $k_z$ (Extended Data Fig. 1), confirming the quasi-two-dimensional nature of the electronic states, consistent with its van der Waals layered structure.

Fermi-surface (FS) mapping in the $k_x$–$k_y$ plane reveals three distinct pockets: two square-like pockets centred at the Γ and M points of the Brillouin zone, and an elliptical pocket at the X point (Fig. 2a,b). The experimental FSs are inconsistent with DFT calculations assuming ferromagnetic and altermagnetic orders (Extended Data Fig. 2). In contrast, they are well reproduced by nonmagnetic DFT calculations (Fig. 2c), indicating the absence of long-range magnetic order in $V_2Se_2O$. This behaviour contrasts sharply with that of $KV_2Se_2O$, where DFT calculation with altermagnetic order shows excellent agreement with ARPES measurements[21].

Band dispersions near the Fermi level ($E_F$) were examined along high-symmetry lines Γ–X, X–M, and Γ–M (Fig. 2d–i). To resolve these bands clearly, we selected different photon energies, light polarizations, and momentum cuts. The α band forms the Γ-centred square-like pocket, except for its corners, which originate from the nearly linear γ band. The α band also generates the M-centred pocket, while the β band produces the X-centred pocket. These dispersions establish that all FS pockets are electron-like. In addition, we identify a hole-like δ band located below –0.3 eV along Γ–M. These results confirm a multiband electronic structure near $E_F$ in $V_2Se_2O$.

## Orbital-selective electron–electron correlation

In Fig. 3c, we compare the experimental dispersions with those from DFT calculations. The β and γ bands generally align with theoretical predictions, except for a clear kink near –60 meV in the β band, indicative of strong coupling to bosonic modes. In contrast, the α band shows significant bandwidth renormalization—by a factor of ~2.0—relative to DFT, signalling substantial electron–electron correlation. Remarkably, the correlation is exclusively confined to the α band, while the β and γ bands remain essentially unaffected. Such extreme selectivity distinguishes $V_2Se_2O$ from typical correlated multiorbital systems, in which correlation usually involves multiple bands near $E_F$[26,31,34], and thus provides a unique opportunity to investigate the microscopic mechanisms driving orbital-dependent correlation phenomena.

To unravel the origin of this selectivity, we analysed the orbital character of the

near-$E_F$ bands using orbital-projected DFT calculations. The results indicate that the $\alpha$ and $\beta$ bands originate mainly from V 3$d$ orbitals, whereas the $\gamma$ band exhibits strong hybridization with Se 4$p$ states (Extended Data Fig. 3). The local environment around each V atom is a distorted Se$_4$O$_2$ octahedron. We therefore define a local coordinate system with the $z$-axis aligned along the O–V–O direction and the $x$–$y$ plane coinciding with the VSe$_4$ plane (Fig. 3a). Within this framework, the orbital contributions become clear: the $\alpha$ band derives from the $d_{xy}$ orbital and the $\beta$ band from the degenerate $d_{xz/yz}$ orbitals (Fig. 3b). The correlation-induced renormalization unique to the $\alpha$ band is thus directly linked to the $d_{xy}$ orbital.

To quantitatively reconcile experiment and theory, we employed the LDA + Gutzwiller method, which incorporates on-site electron correlation beyond standard DFT[58–60]. Applying a uniform Coulomb interaction ($U$) to all V 3$d$ orbitals leads to overall band narrowing, inconsistent with our ARPES data (Extended Data Fig. 4). In contrast, selectively applying $U$ only to the $d_{xy}$ orbital yields substantial renormalization of the $\alpha$ band, while leaving the $\beta$ and $\gamma$ bands nearly unchanged (Extended Data Fig. 5). With $U$ = 7 eV for the $d_{xy}$ orbital, the calculated band structure shows excellent agreement with the experimental dispersions (Fig. 3d).

Further insights come from orbital-resolved density of states and occupancy analyses (Fig. 3e). The $d_{xy}$ orbital is confined to a narrow energy window (~2 eV) near $E_F$ and has an occupancy of $n_e$ = 0.991, placing it very close to half-filling. In comparison, the $d_{xz/yz}$ orbitals extend over a broad energy range (~5 eV) and deviate markedly from half-filling. The two $e_g$ orbitals ($d_{x^2-y^2}$ and $d_{z^2}$) are pushed above $E_F$ by the octahedral crystal field. Within the single-orbital Hubbard picture, strong correlation is favoured under conditions of narrow bandwidth and proximity to half-filling. In V$_2$Se$_2$O, both criteria are satisfied by the $d_{xy}$ orbital, whereas the other 3$d$ orbitals show pronounced deviations. This contrast demonstrates that, even in multiorbital systems, the filling number and the bandwidth remain decisive factors governing the correlation strength of individual orbitals.

Although the bare Coulomb interaction significantly exceeds the $d_{xy}$ bandwidth

(7 eV *vs.* 2 eV), electron correlation only moderately renormalizes the $\alpha$ band and do not drive an orbital-selective Mott transition. This behaviour suggests that screening effects from other partially filled orbitals effectively reduce correlation in the $d_{xy}$ channel[22,50]. This highlights the essential role of interorbital interactions in modulating electron correlation, indicating that the correlation strength is dictated not only by on-site interaction but also by dynamic interorbital screening.

**Orbital-selective electron–phonon coupling**

We next examine the dispersion kink in the band structure—a hallmark of electron–boson coupling. A kink anomaly is clearly observed in the $\beta$ band along both Γ–X and X–M (Fig. 2g,h). Quantitative analysis using Lorentzian fits to momentum distribution curves (MDCs) yields a renormalized Fermi velocity of $v_F = 0.651$ eV·Å along X–M (Fig. 4b,c), substantially lower than the bare velocity $v_F^0 = 2.355$ eV·Å. Within a simple electron–boson coupling model, the coupling constant is estimated as $\lambda = v_F^0/v_F - 1 \approx 2.6$, placing V$_2$Se$_2$O in the strong-coupling regime. Further insights come from the self-energy extracted from the data: the real part (ReΣ) exhibits a broad peak around –60 meV (Fig. 4d), while the imaginary part (ImΣ) increases steadily and saturates below –80 meV (Fig. 4e). These features are consistent under the Kramers−Kronig (K−K) transformation, confirming the reliability of our analysis.

The bosonic modes could in principle be of magnetic or phononic origin. To distinguish between them, we investigated the temperature evolution of the kink. It remains clearly visible from 40 K to 200 K, with only a slight reduction in the magnitude of ReΣ (Extended Data Fig. 6). Moreover, a similar kink is observed in the corresponding band of KV$_2$Se$_2$O—a structurally similar compound with a distinct magnetic order[21]. The persistence of the kink across a wide temperature range and in materials with contrasting magnetic properties strongly disfavours a magnetic origin, instead pointing to a phononic mechanism. This interpretation is supported by electron–phonon coupling calculations. The calculated real part (ReΣ$_{cal}$) shows a broad peak around –60 meV (Fig. 4f), in quantitative agreement with experiment, though the theoretical magnitude is smaller—likely due to the common underestimation of coupling strength in DFT-based approaches.

The calculated phonon spectra separate into two groups: low-energy modes 1–13 distributed below 50 meV and high-energy optical modes 14–15 near 80 meV (Fig. 4g). Both contribute comparably to ReΣ, producing a broad peak around –60 meV (Fig. 4f), indicating that the kink in the $\beta$ band results from coupling to multiple phonon modes. In contrast, the $\alpha$ band shows a much weaker self-energy feature near –35 meV, arising mainly from modes 1–13, while its coupling to modes 14–15 is negligible (Extended Data Fig. 7). The enhancement of the kink in the $\beta$ band is therefore attributed to strong coupling to modes 14–15.

These high-energy modes involve primarily oxygen vibrations within the $a$–$b$ plane (Extended Data Fig. 8), which strongly modulate the crystal potential for orbitals extending toward them. The $d_{xz/yz}$ orbitals are oriented along the V–O direction, resulting in substantial spatial overlap with oxygen displacements and enhanced electron-phonon matrix elements. In contrast, the $d_{xy}$ orbital lies within the VSe$_4$ plane, spatially separated from oxygen sites, and thus couples weakly to these modes. This geometric distinction naturally accounts for the enhancement of electron–phonon coupling in the $\beta$ band, highlighting the microscopic mechanism behind orbital-selective electron-phonon interactions in V$_2$Se$_2$O.

## Summary and Outlook

Our study reveals an exceptional example of orbital-selective many-body interactions, in which electron–electron correlation and electron–phonon coupling manifest selectively in distinct orbital-derived bands. This pronounced orbital differentiation pinpoints the microscopic origins of orbital selectivity: correlation-induced renormalization arises from narrow bandwidth and proximity to half-filling, while orbital-dependent electron–phonon coupling is governed by spatial overlap between orbital wavefunctions and atomic vibrations. Further control through external pressure, doping, or symmetry breaking could enable precise tuning of these orbital-dependent interactions, providing a pathway to realize exotic quantum phases—such as orbital-selective superconductivity or altermagnetism—mediated by anisotropic many-body effects.

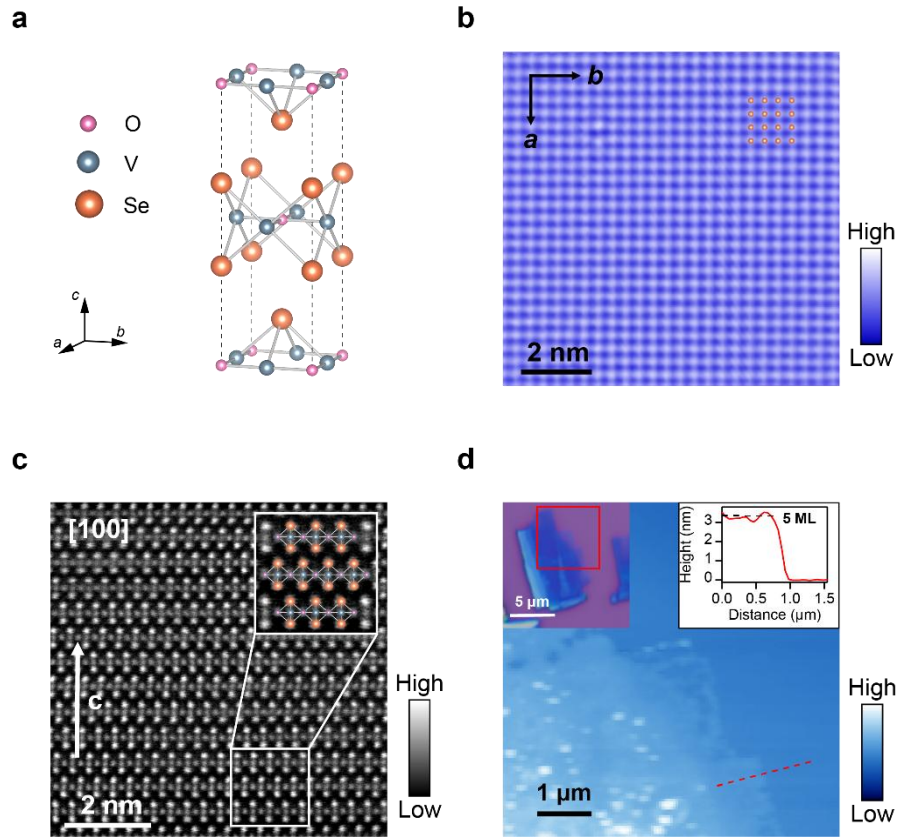

**Fig. 1. Crystal structure and exfoliation. a**, Crystal structure of $V_2Se_2O$. **b**, Atomically resolved STM image (sample bias $V_s = -100$ mV, tunnelling current $I_t = 300$ pA) of the cleaved (001) surface, overlaid with the Se atomic lattice model (yellow spheres). The topographic image was acquired in constant-current mode at 4.8 K. **c**, Atomic-resolution STEM image viewed along the [010] zone axis. Inset, magnified region from the white box, with coloured spheres indicating the atomic positions corresponding to the intensity spots. **d**, Atomic force microscopy image of a mechanically exfoliated flake, corresponding to the region marked by the red box in the optical microscopy image (top-left inset). Top-right inset, height profile along the red dashed line in **d**, consistent with the step height of a five-monolayer $V_2Se_2O$ flake (dashed curve).

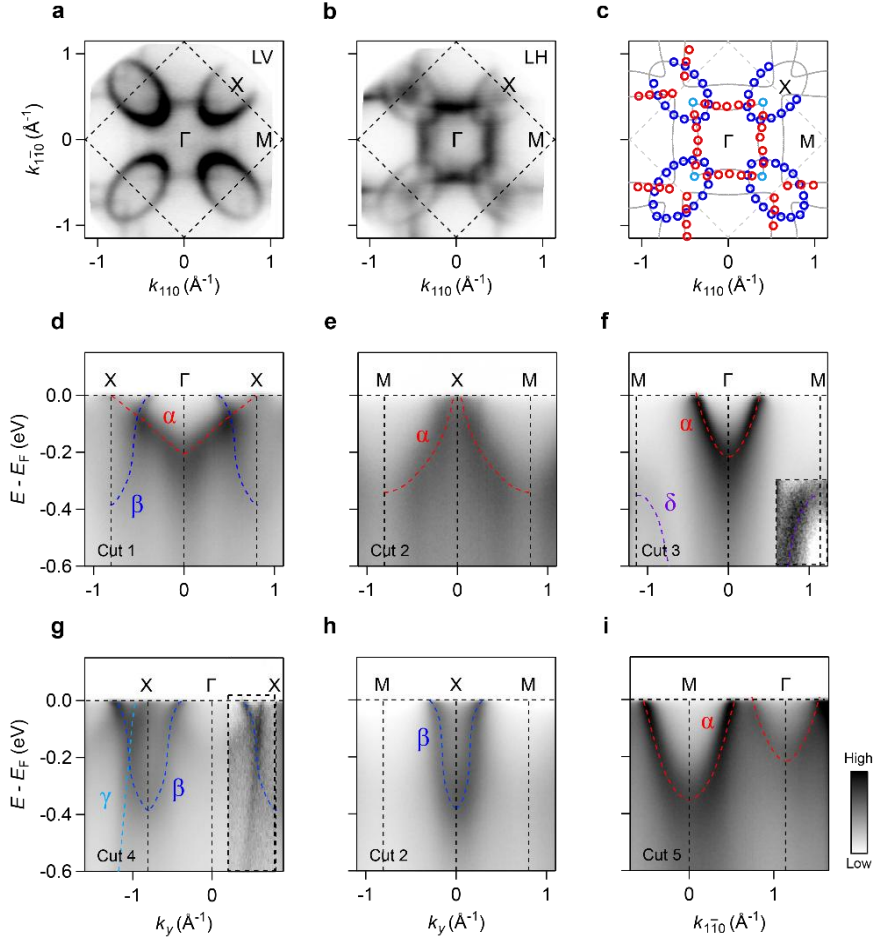

**FIG. 2. FSs and band dispersions. a,b**, ARPES intensity maps at $E_F$ measured with photon energy $h\nu$ = 68 eV under linear vertical (LV) and linear horizontal (LH) polarizations. **c**, FSs extracted from **a** and **b** (open circles) compared with DFT calculations of monolayer $V_2Se_2O$ in the nonmagnetic state (grey lines). Red, blue and light blue colours denote the Fermi crossing positions of the $\alpha$, $\beta$ and $\gamma$ bands, respectively. Dashed lines in **a**–**c** mark the Brillouin zone boundary of monolayer $V_2Se_2O$. **d,g**, ARPES intensity plots along Γ–X under LH polarization with $h\nu$ = 84 and 68 eV. Data within the dashed box in **g** were collected with $h\nu$ = 40 eV. **e,h**, ARPES intensity plots along X–M under LH and LV polarizations with $h\nu$ = 68 eV. **f,i**, ARPES intensity plots along Γ–M under LH polarization with $h\nu$ = 84 and 64 eV. Data within the dashed box in **f** are shown with adjusted colour scale to highlight the $\delta$ band. Red, blue, light blue and purple dashed lines in **d**−**i** are guides to the $\alpha$, $\beta$, $\gamma$ and $\delta$ band dispersions, respectively. The momentum positions of the cuts in **d**−**i** are shown in Extended Data Fig. 1a. All ARPES data were collected at 25 K.

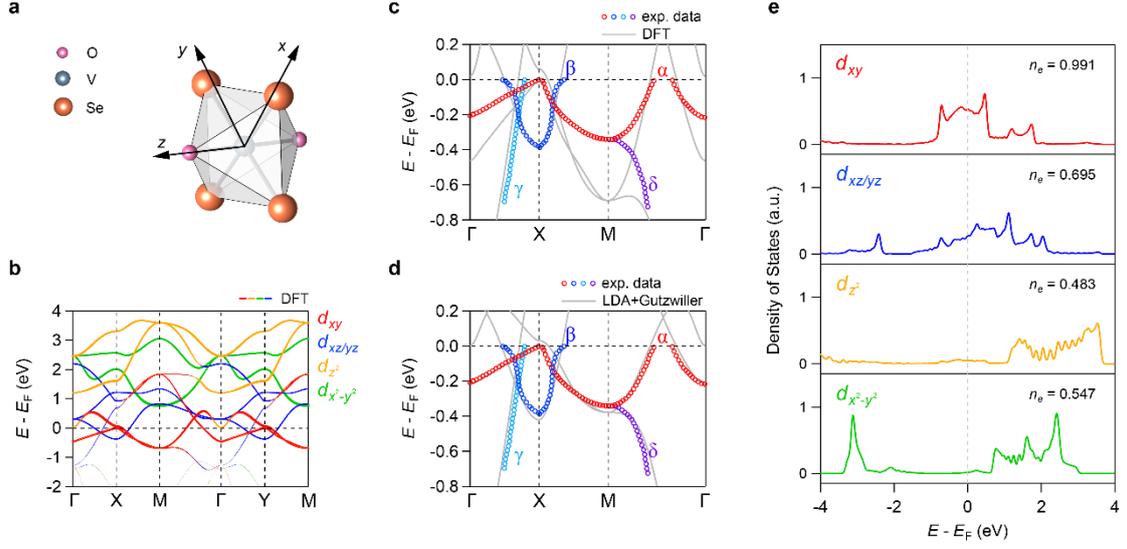

**FIG. 3. Orbital-projected calculations and bandwidth renormalization. a**, Schematic of the local coordinate system for a V-centred $Se_4O_2$ octahedron. The *z*-axis is along the V–O bonds, while the *x*- and *y*-axes are approximately along the V–Se bonds. **b**, Orbital-projected band structure of monolayer $V_2Se_2O$ from DFT. The $d_{xy}$, $d_{xz/yz}$, $d_{z^2}$ and $d_{x^2-y^2}$ orbitals are shown in red, blue, orange and green, respectively, with line thickness proportional to weight of projected orbital. **c**, Comparison between experimental dispersions (open circles) and DFT calculations (grey lines). **d**, Comparison between experimental dispersions (open circles) and LDA + Gutzwiller calculations with $U = 7$ eV applied to the $d_{xy}$ orbital (grey lines). **e**, Orbital-projected density of states and occupancy $n_e$ of V 3*d* orbitals obtained from DFT.

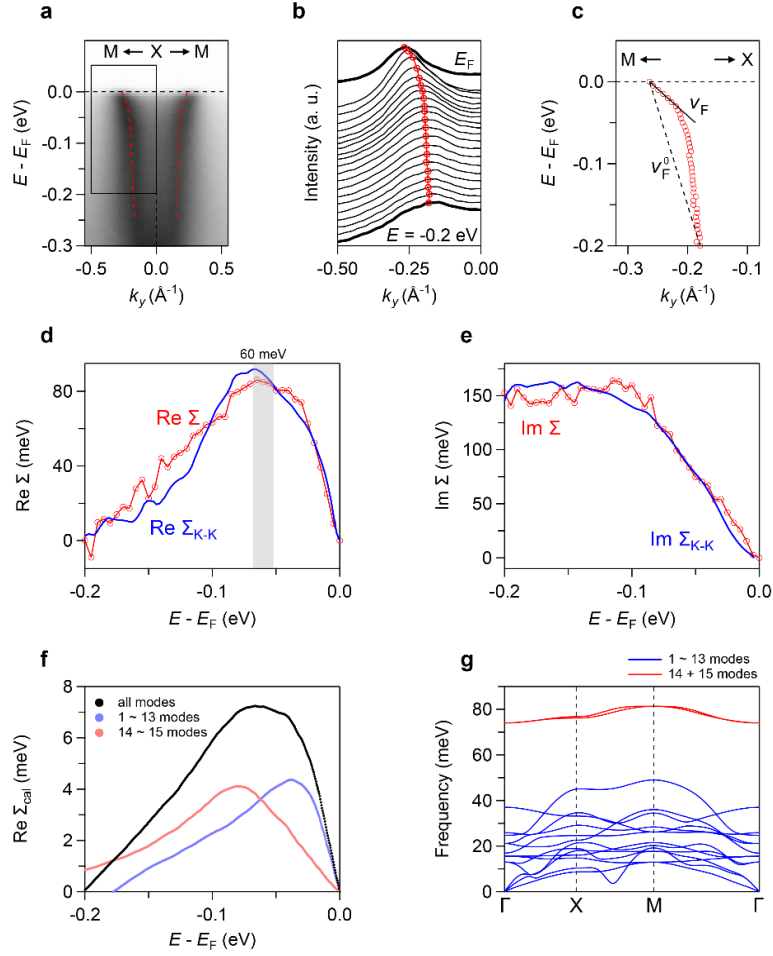

**FIG. 4. Dispersion kink and phonon spectra. a**, ARPES intensity plot measured with $h\nu$ = 68 eV under LV polarization, showing the $\beta$ band along X−M. Red dashed curves mark the extracted dispersion. **b**, MDCs of the data within the black box in **a**. Red circles indicate peak positions from Lorentzian fits. **c**, Extracted band dispersion with solid and dashed lines denoting renormalized and bare bands, respectively. **d**, ReΣ (red circles) compared with the K−K transformed ReΣ$_{K-K}$ (blue curve) obtained from ImΣ in **e**. **e**, ImΣ (red circles) compared with the K−K transformed ImΣ$_{K-K}$ (blue curve) derived from ReΣ in **d**. Background contributions from electron–impurity scattering (constant) and electron–electron correlation ($\propto$ binding energy$^2$) were subtracted. **f**, Calculated ReΣ$_{cal}$ for the $\beta$ band from electron–phonon coupling. Black, blue and red dots denote contributions from all phonon modes, modes 1–13, and modes 14–15, respectively. **g**, Phonon dispersions calculated by density functional perturbation theory. Blue and red lines correspond to modes 1–13 and 14–15, respectively.

**Methods**

**Sample synthesis and characterization.** $V_2Se_2O$ polycrystals and single crystals were obtained from $KV_2Se_2O$ polycrystals and single crystals via a soft-chemical route with deintercalation of $K^+$. The precursor $KV_2Se_2O$ was synthesized as previously reported from K (2N, lump), $V_2O_5$ (3N, powder), V (3N, powder), and $VSe_2$ by solid-state reaction for polycrystals, and using KSe as the self-flux agent for single crystals[61]. The as-prepared $KV_2Se_2O$ samples were immersed in excessive deionized water at room temperature for two weeks, then washed several times with deionized water and dried under nitrogen atmosphere at 60 °C to remove residual water. The obtained black polycrystals and fragile single crystals were air-stable.

The crystal structure of $V_2Se_2O$ was characterized by X-ray diffraction (XRD) using a XtaLAB Synergy Diffractometer with Mo Kα radiation at room temperature. Atomic-resolution high-angle annular dark-field scanning transmission electron microscopy (HAADF-STEM) was performed using a JEOL ARM200F microscope equipped with probe- and image-side spherical aberration correctors, operating at 200 kV. HAADF images were acquired using an annular detector with semi-collection angles of 90–370 mrad, achieving a nominal probe resolution of 78 pm. The intensities of atomic columns in HAADF images scale approximately $Z^{1.7}$, where Z is the atomic number of the constituent elements[62,63]. Cross-sectional specimens oriented along the [100] zone axis were fabricated via focused ion beam (FIB). Crystal flakes were mechanically exfoliated[64] onto $Si/SiO_2$ substrates (300 nm). The substrates were pre-cleaned by oxygen plasma (50 sccm, 100 W, 100 mtorr) to enhance interfacial van der Waals forces, thereby facilitating the subsequent thinning process.

**Angle-resolved photoemission spectroscopy.** All ARPES measurements were carried out at the "Dreamline" (BL09U) beamline of the Shanghai Synchrotron Radiation Facility (SSRF) using a Scienta Omicron DA30 electron analyser under both linear vertical (LV) and linear horizontal (LH) polarizations. Data were collected over a photon energy range of 25–90 eV with a beam spot size of 30 × 50 μm. The samples were cleaved in situ at 25 K under a base pressure better than $3 \times 10^{-11}$ mbar.

To analyse the dispersion kink, we extracted the complex electron self-energy $\Sigma = \mathrm{Re}\Sigma + i\mathrm{Im}\Sigma$. $\mathrm{Re}\Sigma$ was obtained by subtracting a linear bare-band dispersion from the experimentally determined quasiparticle dispersion. This bare band was determined by linearly fitting the dispersion in the high-binding-energy region (–250 to –200 meV), where interaction effects are assumed to be negligible. $\mathrm{Im}\Sigma$ was calculated from the Lorentzian full width at half maximum (FWHM) of the MDCs, $\Gamma_{\mathrm{MDC}}$, using the relation $\mathrm{Im}\Sigma = |v_F^0| \times \Gamma_{\mathrm{MDC}} / 2$, where $v_F^0$ is the bare-band velocity obtained from the linear fit. To isolate the scattering channel associated with the kink, we subtracted a background consisting of a constant term (electron–impurity scattering) and a quadratic term in energy (electron–electron correlation).

**Calculations.** First-principles calculations were carried out using the Vienna ab initio simulation package (VASP)[58] within the generalized gradient approximation of the Perdew–Burke–Ernzerhof (PBE) type[59] for the exchange-correlation functional. Spin–orbit coupling (SOC) was not included, as its influence on the band structure is negligible. Since the electronic dispersion along the out-of-plane direction is very weak, calculations were performed on a monolayer $V_2Se_2O$ slab with a vacuum spacing of 30 Å along the $z$-axis. For self-consistent field (SCF) calculations, we employed a Monkhorst–Pack $k$-point mesh of $7 \times 7 \times 3$ and a plane-wave energy cutoff of 600 eV. To construct the tight-binding Hamiltonian, we used the Wannier90 package[65,66] to obtain maximally localized Wannier functions of V $d$, Se $p$ and O $p$ orbitals. To include strong electronic correlation, LDA + Gutzwiller calculations were performed using the RTGW2020 package[60], with an on-site Coulomb interaction parameter of $U = 7$ eV applied to the V $d_{xy}$ orbitals.

Phonon spectra were calculated using density functional perturbation theory (DFPT) as implemented in the Quantum ESPRESSO package[67]. The structure was fully relaxed with a force convergence threshold of $10^{-3}$ eV·Å$^{-1}$. Calculations employed Optimized Norm-Conserving Vanderbilt (ONCV) pseudopotentials[68], a kinetic energy cutoff of 130 Ry for the plane-wave basis, and Fermi–Dirac smearing of 0.01 Ry. Self-consistent calculations used an $11 \times 11 \times 1$ Monkhorst–Pack $k$-point mesh with a total energy convergence threshold of $10^{-13}$ Ry. Dynamical matrices were computed on a $5 \times 5 \times 1$ $q$-point mesh and interpolated to obtain the phonon dispersions. Electron–phonon coupling (EPC) and its impact on the electronic self-energy were evaluated using the EPW code[69]. The calculations were performed at 10 K, with a delta-function smearing of 0.03 Ry for energy conservation. Brillouin-zone integration over phonon momentum was carried out on a dense $100 \times 100 \times 1$ $q$-point mesh.

**Data availability**

The data that support the findings of this study are available from the corresponding authors on reasonable request.

**Acknowledgement**

We thank Xi Dai, Tiantian Zhang for fruitful discussions. We thank Yaobo Huang for assistance with the ARPES experiments at the SSRF. This work was supported by the Ministry of Science and Technology of China (2022YFA1403800, 2022YFA1403903, and 2023YFA1406500), the National Natural Science Foundation of China (12525409, U22A6005, 12274440, T2325028, 12134019, 12274459, and 62488201). The authors acknowledge the BL09U beamline at the Shanghai Synchrotron Radiation Facility (SSRF) and the A8 experimental station at the Synergetic Extreme Condition User Facility (SECUF). H.W. acknowledges support from the New Cornerstone Science Foundation through the XPLORER PRIZE. H.L. acknowledges support by the CAS Pioneer Hundred Talents Program.

**Author contributions**

H.L., T.Q., and H.W. supervised the project. M.H., G.Q., T.Q. and H.L. performed the ARPES experiments; J.C., Y.H., Z.T., H.L., and G.C. synthesized the single crystals; Z.S., Z.L. and H.W. performed the calculations; M.H. and Z.W. performed the STM experiments; X.M. and H.Y. performed the TEM experiments; J.Z. and G.Z. exfoliated the sample into thin flakes; D.T. and L.C. performed the AFM experiments; M.H., Z.S., H.W. T.Q., and H.L. plotted the figures and wrote the manuscript with contributions from all authors.

**Competing interests**

The authors declare no competing interests.

**Extended Data Table 1 Crystallographic data of $V_2Se_2O$ obtained from single crystal X-ray diffraction refinement at room temperature.**

| | Chemical formula | | | $V_2Se_2O$ | |
|---|---|---|---|---|---|
| | Space group | | | *I4/mmm* | |
| | *a* (Å) | | | 3.8962(9) | |
| | *c* (Å) | | | 11.712(7) | |
| | *V* (Å$^3$) | | | 177.79(13) | |
| Atom | Site | *x* | *y* | *z* | $U_{iso}$ (Å$^2$) |
| V | 4c | 0 | 0.5 | 0.5 | 0.015 |
| Se | 4e | 0.5 | 0.5 | 0.3527(3) | 0.017 |
| O | 2b | 0 | 0 | 0.5 | 0.012 |

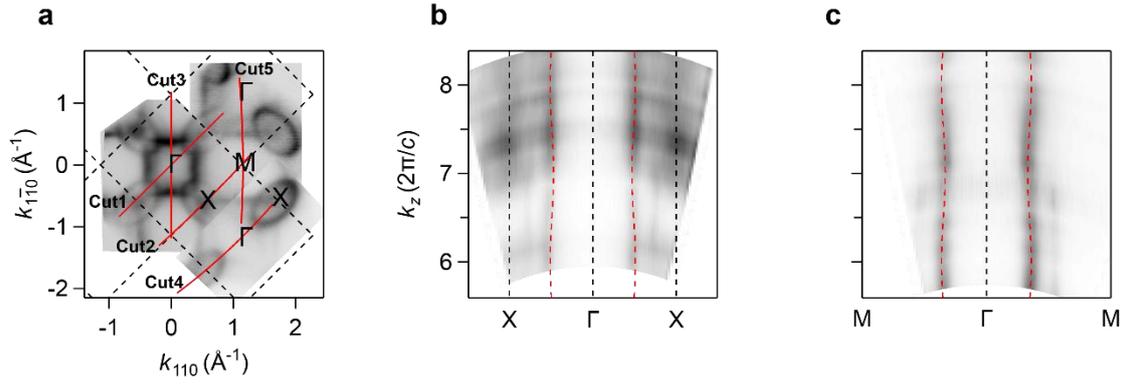

**Extended Data Fig. 1 | In-plane and out-of-plane FS mapping. a**, ARPES intensity map at $E_F$ in the $k_x$–$k_y$ plane. Red lines indicate the momentum cuts shown in Fig. 2d–i. **b,c**, ARPES intensity maps at $E_F$ along Γ–X and Γ–M measured with varying photon energies. Red dashed lines indicate weak $k_z$ dispersion with a periodicity of $4\pi/c$.

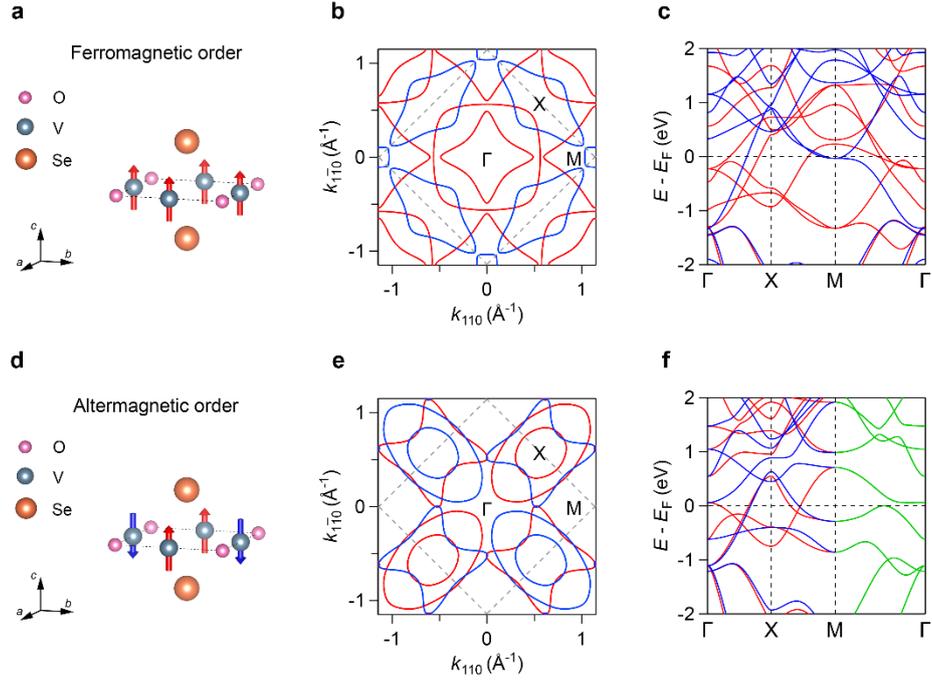

**Extended Data Fig. 2 | Calculated spin-resolved FSs and band structures along high-symmetry lines for different assumed magnetic orders. a–c**, Schematic magnetic configuration, FSs, and band structure for the ferromagnetic order. **d–f**, Same as **a–c** but for the altermagnetic order. Red, blue and green curves denote spin-up, spin-down and spin-degenerate bands, respectively.

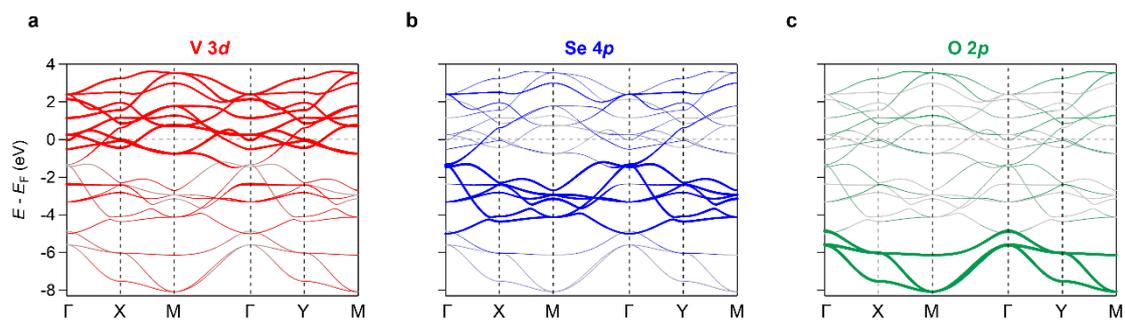

**Extended Data Fig. 3 | Calculated orbital-resolved band structures along high-symmetry lines. a**–**c**, Band structures projected onto the orbitals of V, Se, and O atoms, with line thickness proportional to orbital weight.

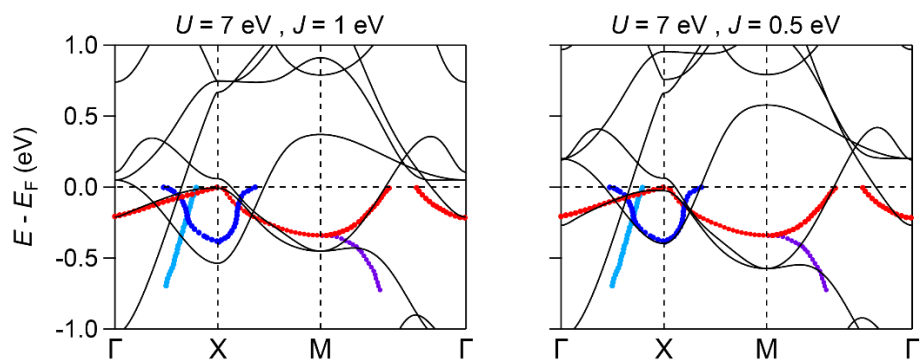

**Extended Data Fig. 4 | LDA + Gutzwiller calculations with *U* and Hund's coupling (*J*) applied to all V 3*d* orbitals.** Results for $U = 7$ eV with $J = 1$ and 0.5 eV are shown as representative cases. These parameters provide comparatively better agreement with experiment than other tested values. Nevertheless, the calculated bands show overall narrowing and fail to capture the pronounced orbital selectivity observed experimentally. Solid circles denote experimental band dispersions.

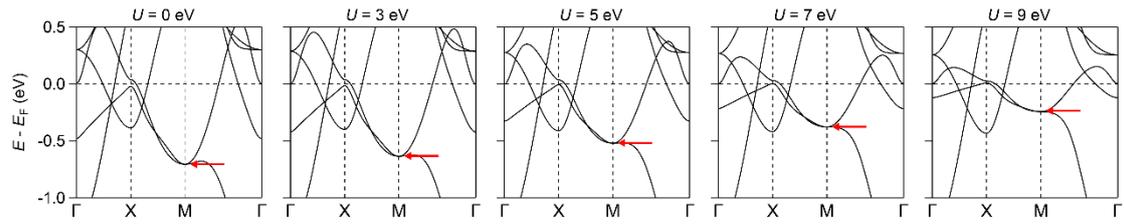

**Extended Data Fig. 5 | LDA + Gutzwiller calculations with *U* applied only to the $d_{xy}$ orbital.** With increasing *U*, the $d_{xy}$-derived band becomes progressively narrower (red arrow), while the other bands remain nearly unchanged. The case of $U = 7$ eV gives the closest agreement with ARPES measurements.

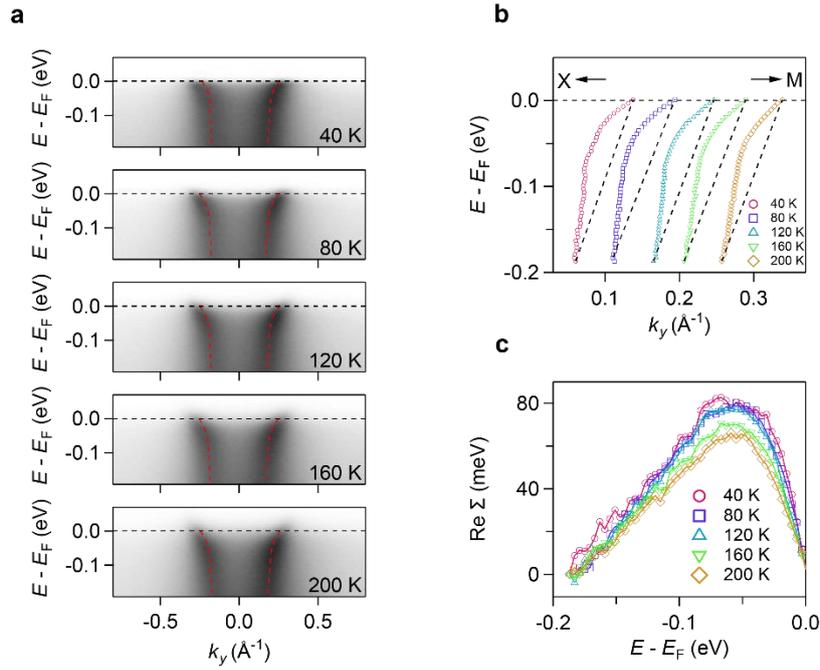

**Extended Data Fig. 6 | Temperature dependence of the dispersion kink. a**, ARPES intensity plots measured along X–M under LV polarization with $h\nu = 68$ eV at different temperatures. Red dashed curves are extracted band dispersions from Lorentzian fits to MDCs. **b**, Band dispersions extracted from **a**. Dashed lines indicate the bare band. The dispersions are offset horizontally for clarity. **c**, Re$\Sigma$ at different temperatures.

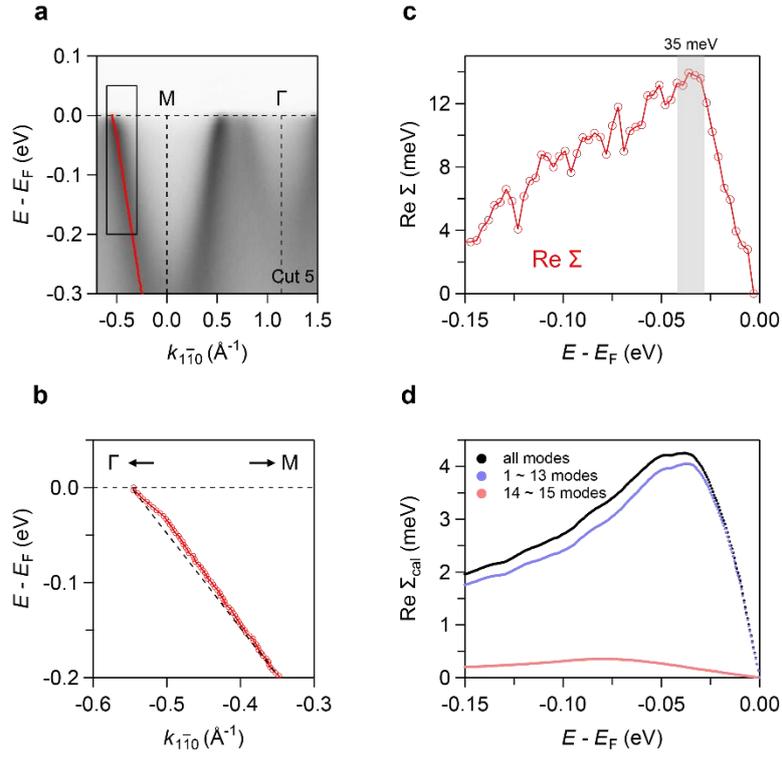

**Extended Data Fig. 7 | Dispersion kink of the *α* band. a,** ARPES intensity plot measured with *hv* = 64 eV under LH polarization, showing the *α* band along Γ−M. Red dots are dispersion extracted from Lorentzian fits to MDCs. **b,** Extracted dispersion within the black box in **a.** Black dashed line indicates the bare band. **c,** ReΣ for the *α* band. **d,** ReΣ$_{cal}$ for the *α* band from electron–phonon coupling. Black, blue, and red dots denote contributions from all phonon modes, modes 1–13, and modes 14–15, respectively.

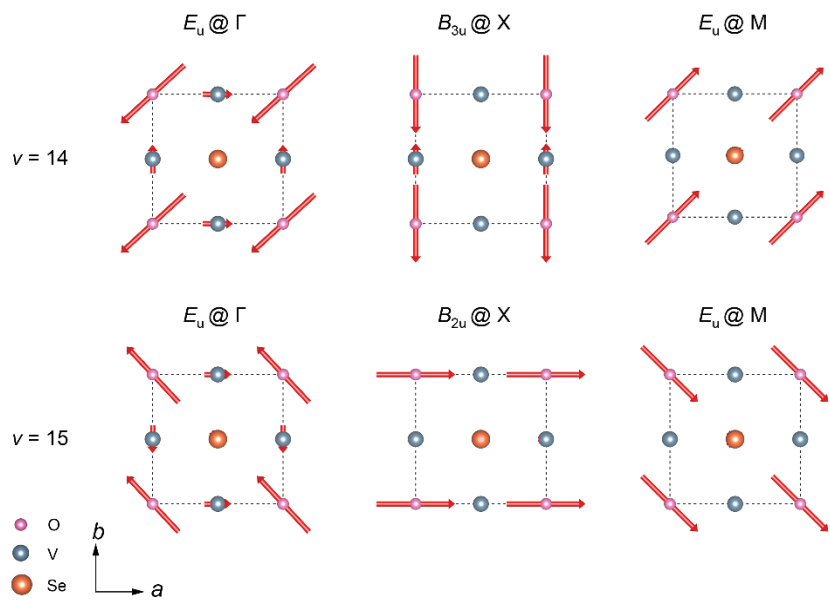

**Extended Data Fig. 8 | In-plane atomic displacements for phonon modes 14 and 15 at the Γ, X and M points.** The arrow length represents the vibration amplitude.